\documentstyle[12pt]{article}
\begin{document}
\begin{center}
{\Large \bf
THE APPLICABILITY OF THE EQUIVALENCE THEOREM IN  $\chi PT$}
\vskip 2.0cm
{\large \ Antonio DOBADO \footnote{E-mail: dobado@cernvm.cern.ch}
and Jos\'e Ram\'on PELAEZ \footnote{E-mail: pelaez@fis.ucm.es} }  \\
\vskip 0.5cm
 Departamento de F\'{\i}sica Te\'orica  \\
 Universidad Complutense de Madrid\\
 28040 Madrid, Spain \\
\vskip 0.5cm
and
\vskip 0.5cm
{\large \ Mar\'{\i}a Teresa URDIALES \footnote{E-mail: mayte@ccuam3.sdi.uam.es}
\\
\vskip 0.5cm}
 Departamento de F\'{\i}sica Te\'orica  \\
 Universidad Aut\'onoma de Madrid\\
 28049 Madrid, Spain \\
 \vskip .5cm
{\bf \large  Contributed paper to the $27th$ International
Conference in High Energy Physics}
\vskip 0.5cm
\begin{abstract}
We have explicitely
 calculated the tree level elastic scattering cross sections of two
longitudinal
gauge bosons, up to four derivatives in the
 chiral expansion both with and without using the Equivalence Theorem (ET).
The numerical results  show the existence of new and severe restrictions
in the ET energy applicability range,
as it was stated in our recent derivation, which we also review here, of the
 precise ET version in the Chiral Lagrangian description of the
Standard Model Symmetry Breaking Sector.
\end{abstract}
\end{center}
\newpage

\section{Introduction}

In this paper we try to clarify the problem of the applicability of
the so-called Equivalence Theorem (ET)	\cite{ETfirst,ChMKG,Gou} which relates,
at
high energies, the longitudinal
electroweak gauge bosons $S$ matrix elements with those elements where these
gauge bosons have been replaced by their corresponding would be Goldstone
Bosons
(GB). This relation is very useful to obtain
information from the future LHC data about the Standard Model (SM)
Symmetry Breaking Sector (SBS), since computations are
much easier to do for scalars than for longitudinal gauge bosons.

 Despite the very precise data	collected at LEP we almost have no
information on the SMSBS and we do not know, at present,
which is the dynamics responsible
for the spontaneous breaking of the electroweak group $SU(2)_L\times U(1)_Y$
to  the electromagnetic group $U(1)_{em}$, so it would be interesting to
develop
a model independent framework to describe phenomenologically the
SBS mechanism. Recently this approach has been followed  borrowing
a formalism from low-energy hadron physics that is called Chiral
Perturbation Theory ($\chi$PT) \cite{Wein,DoHe}, and it has been proved to be
also quite
useful for the analysis of the precision measurements  obtained at LEP
\cite{DoEs}.

In order to apply $\chi$PT to the SMSBS description one
assumes that there must be some  physical system coupled to the SM with a
global
symmetry breaking from a gauge group $G$ to another gauge group $H$ producing
the
spontaneous symmetry breaking of $SU(2)_L\times U(1)_Y$ to
$U(1)_{em}$ which yields the $W^{\pm}$ and $Z$ masses through
the standard Higgs mechanism. The  GB related to the global $G$ to $H$ symmetry
breaking are nothing but coordinates in the coset space $G/H$ and their low
energy dynamics is described by a gauged Non-Linear Sigma Model plus an
infinite
number of higher derivative terms (but finite for practical purposes) needed
for the renormalization of the model.

As we have said before both $\chi $PT and the ET have been used together to
describe
the scattering of  gauge bosons longitudinal components, even though  a
rigorous proof of this theorem in the Chiral Lagrangian description of the SM
has only
been presented very recently \cite{DoPe}, and although it is known that
correction factors have to be taken into account due to the different
renormalization
of the GB and the gauge bosons even in the simple original formulation of the
Minimal
Standard Model (MSM) \cite{Yao}, as well as in the Effective Lagrangian
formalism \cite{He}.

In this paper we follow the proof of the precise statement of the theorem
valid for a chiral lagrangian description of the SMSBS including the
renormalization
effects, which was given in \cite{DoPe}. That derivation is based on the nice
formal proof
of the ET for the MSM by Chanowitz and Gaillard \cite{ChMKG}, and consists in
obtaining
Ward-Slavnov Taylor identities from the BRS symmetry \cite{BRS} of the model
and then to translate these relations between Green functions, to $S$ matrix
elements
but we take into account  the peculiarities  of $\chi $PT and include
renormalization factors.
To implement the BRS symmetry we could have followed the standard Faddeev-Popov
quantization procedure, but since we want a proof valid for any GB
parametrization
choice, we will use a more general method given in
 \cite{Baulieu} that deals more elegantly with non-linear gauge fixing
conditions which will be needed
to ensure the covariance of the quantum Lagrangian under coordinate changes in
the GB
coset.

Once we have obtained the precise formulation of the ET in the Chiral
Lagrangian
formalism, we will discuss its energy applicability range coming from the
different
approximations that we need in its derivation. We will also show some numerical
results for the  elastic scattering amplitude of two longitudinal gauge bosons,
which will
allow us to compare the tree level computations of the chiral lagrangian up to
four derivatives
, both with or without using the ET, and therefore to check the applicability
window
obtained in the proof presented in \cite{DoPe}

\section{The chiral lagrangian formalism and the SMSBS}

 The known facts on the SMSBS impose some conditions on  $G$ and $H$:

 a) $dimK=dim G/H=3$ since we need three GB to give mass to  $W^{\pm}$ and $Z$.

b) $G$ contains $SU(2)_L\times U(1)_Y$ so that the symmetry breaking sector
couples
 to the electroweak gauge bosons.

c) $H$ contains the custodial group $SU(2)_{L+R}$ in order to ensure the
experimental
relation $\rho \simeq 1$ \cite{Sikivie}. This constraint yields $\rho =1$ once
the gauge couplings are set to zero and implies that the photon is massless
since $U(1)_{em}$ is
contained in $SU(2)_{L+R}$.

 It has been shown in \cite{DoPe} that these conditions completely determine
the $G$ and
$H$ groups to be $G=SU(2)_R\times SU(2)_L$ and $H=SU(2)_{L+R}$ and thus
$K=G/H=S^3$. Therefore, we can describe the SMSBS in the Chiral Lagrangian
formalism
 as a gauged non-linear sigma model based on the coset space
$K=G/H=SU(2)_L\times
SU(2)_R/SU(2)_{L+R}=S^3$ with gauge group  $SU(2)_L\times U(1)_Y$. Thus we can
write the lagrangian
\begin{eqnarray}
 {\cal L}_g    &=&{\cal L}_{YM}^L+{\cal L}_{YM}^Y
+\frac{1}{2}g_{\alpha\beta}(\omega)D_{\mu}\omega^{\alpha}
D^{\mu}\omega^{\beta} \\ \nonumber
 &+& higher \; covariant \; derivative \; terms
\end{eqnarray}
where ${\cal L}_{YM}^L$ and ${\cal L}_{YM}^Y$ are the usual
Yang-Mills lagrangians for the
 $SU(2)_L$ and $U(1)_Y$ gauge fields $W_{\mu}^a$ and $B_{\mu}$; the
$\omega^{\alpha}$
 fields are arbitrary coordinates on the coset $S^3$ chosen so that
  for the classical vacuum $\omega^ \alpha=0$. The non-linear transformation of
the GB
$\omega^{\alpha}(x) $ under the action of an infinitesimal $G$ element defines
the killing vectors $\xi^{\alpha}_{\;a}$
 through $\delta \omega^{\alpha}=\xi^{\alpha}_{\;a}(\omega)$.
Notice that the $a$ index
runs from $1$ to $6$ where the values $1$ to $3$ correspond to
the unbroken $H=SU(2)_L\times SU(2)_R$ generators. Now we can build the  $S^3$
metrics $g_{\alpha\beta}$ through the dreibein
  $e_a=e^{\alpha}_{\;a} \partial/\partial \omega^{\alpha}$
with $e^{\alpha}_{\;a}=\xi^{\alpha}_{\;a+3}$ for $a=1,2,3$ which is nothing but
the set of killing vectors corresponding to the $3$ broken
generators. We define the metrics as $g^{\alpha\beta}=e^{\alpha}_{\;a}e^{\beta
a}$.This procedure ensures that $G$ is the isometry group of $S^3$ with that
metrics.
The covariant derivatives are defined as:
\begin{equation}
D_{\mu}\omega^{\alpha}=\partial_{\mu}\omega^{\alpha}-gl^{\alpha}_{\;a}W_{\mu}^a-g'y^{\alpha}B_{\mu}
\end{equation}
where $l^{\alpha}_{\;a}$ and $y^{\alpha}$ are the killing vectors corresponding
to the gauge groups $SU(2)_L$ and $U(1)_Y$ respectively. The {\it higher
derivative terms}
include
any covariant (in the space-time and the $S^3$ sense) piece containing a bigger
number of covariant derivatives with arbitrary couplings so that
we can reproduce any dynamics compatible with the  $SU(2)_L\times
SU(2)_R/SU(2)_{L+R}=S^3$  to
$SU(2)_L\times U(1)_Y$ symmetry breaking. Thus the gauge transformations are:
\begin{eqnarray}
\delta\omega^{\alpha}&=&l^{\alpha}_{\;a}\epsilon^a_L(x)+y^{\alpha}\epsilon_Y(x)
\\ \nonumber
 \delta W_{\mu}^a&=&
\frac{1}{g}\partial_{\mu}\epsilon^a_L(x)+\epsilon_{abc}
W_{\mu}^b\epsilon_{Lc}(x) \\ \nonumber \delta
B_{\mu}&=&\frac{1}{g'}\partial_{\mu}\epsilon_Y(x)
\end{eqnarray}

\section{The quantum lagrangian and BRS invariance}

The formal proofs of the ET are based in the BRS symmetry of the quantized
lagrangian
\cite{ChMKG,Gou} and they are performed in t'Hooft or renormalizable
($R_{\xi}$) gauges.
 When dealing with the Chiral Lagrangian formalism, the usual linear choice for
the t'Hooft
gauge fixing function does not yield a covariant lagrangian under
reparametrizations
of the GB coset, due to the fact that the GB fields are coordinates,
and their contraction does not transform properly. In the $\chi PT$
applications many
different GB parametrizations are commonly used, that is why we are interested
in an ET proof	valid for any coordinate choice, and therefore, if we want to
use a
t'Hooft gauge fixing function, the dependence on the GB fields should be
nonlinear.
However, it is well known that nonlinear gauge fixing conditions lead to
appearance of
quartic ghost interactions even when they were not present in the original
lagrangian.
That is why we are going now to quantize the model built in the
preceding section following a different procedure \cite{Baulieu}  which deals
more
elegantly with these nonlinear gauge fixing conditions and quartic ghost
interactions.

The above gauge transformations satisfy the Jacobi identity as well as closure
relations (see \cite{DoPe} for details) and therefore we can build the
corresponding
nilpotent ($\bar s$)- s (anti)-BRS transformations
by introducing the anti-commuting ghost fields $c_a$ and  $\bar c_a$,
and the commuting auxiliary field $B_a$  with $a=1,2,3,4$.

For further convenience we will unify the notation so that the first three
values of
the gauge indices $a=1,2,3$ refer to the $SU(2)_L$ group and $a=4$ refers to
$U(1)_Y$, thus the gauge
field $W_{\mu}^a$ with $a=1,2,3,4$ will be defined as $W_{\mu}^a=W_{\mu}^a$
for
$a=1,2,3$ and $W_{\mu}^4=B_{\mu}$. In addition
we introduce the Killing vector $L^{\alpha}_{\;a}$
with $a=1,2,3,4$ as $L^{\alpha}_{\;a}=gl^{\alpha}_{\;a}$ for $a=1,2,3$ and
$L^{\alpha}_{\;4}=g'y^{\alpha}$ and the completely antisymmetric
symbols $f_{abc}$ with $a=1,2,3,4$ as
 $f_{abc}=g\epsilon_{abc}$ for $a=1,2,3$ and $f_{ab4}=0$.

 The  nilpotency  properties (which are equivalent to the Jacobi and closure
relations)
 $ s^2=s \bar s+\bar s s=\bar s ^2=0 $
allow us to define a (anti)-BRS invariant quantum lagrangian as follows:
\begin{equation}
{\cal L}_Q={\cal
L}_g+\frac{1}{2}s \bar s[W_{\mu}^aW^{\mu a}+2\xi f(\omega)+\xi c^a\bar c_a]
\end{equation}
where $f$ is any scalar analytical function with $\partial f(\omega)/\partial
\omega^{\alpha}=\omega^{\alpha}
+O(\omega^2)$. The new terms added to the lagrangian
(see \cite{DoPe} for details) are a generalization of the
 Faddeev-Popov terms in a  t'Hooft like gauges,
which have two main advantages: First, they provide us with a well defined
$R_{\xi}$
propagator to be used in perturbation theory. Second, they cancel the
GB  and gauge boson mixing terms in the lagrangian. In addition, this
generalized method produces other GB-gauge boson and ghost-gauge boson
interactions. For gauges different from that of Landau ($\xi=0$) we also have
quartic ghost interactions and GB-ghosts interactions.

Once we have a (anti)-BRS invariant lagrangian we can derive, using the
standard
functional methods, the Ward-Slavnov-Taylor identities for dimensionally
regularized Green functions. It is worth mentioning that we use
dimensional regularization not only to preserve the (anti)-BRS
invariance but also to avoid the $-\frac{i}{2}\delta^n(0)tr\; \log g$ term that
 appears in the quantum lagrangian of the non-linear sigma model (NLSM)
coming from the path integral measure of the GB fields \cite{delta} .

As a matter of fact, and in order to make
physical predictions, we are interested in renormalized Green functions.
Therefore we have
to consider the renormalized lagrangian which consists on that of eq.4	plus
other terms with the corresponding couplings needed to reproduce all the
divergent structures which appear in the Green functions. At present, the form
of these counterterms is known only up to four derivatives \cite{App}, but
they should also be (anti)-BRS invariant, since if this was not the case,
 the gauge invariance of the model would be anomalous, i.e.,
broken by quantum effects. Nevertheless, it is well known that even
though we had chiral fermions coupled to GB and gauge bosons, the
SM hypercharge assignments are such that
all possible gauge and	mixed gauge-gravitational anomalies, including the
non-perturbative $SU(2)$ discovered by Witten \cite{Witten}, do cancel
when the number of colors is $N_c=3$. As we are interested in reparametrization
invariance we should also worry about the  potential anomalies that could break
 the invariance under coordinate
changes in coset, but these are absent since our NLSM is
defined in a space of lower dimension than space-time, as it was shown in
\cite{Gaume}.
Recently some papers have appeared  where the applicability of the ET is
discussed in models where some anomalies are present, but this is not our case
and we will
only refer the reader to the literature \cite{Don}.

Once we have taken into account all these considerations, we obtain  a
(anti)-BRS invariant lagrangian with  infinite terms which
can be understood as the renormalized lagrangian for a theory with  infinite
couplings written in terms of the bare quantities. However, we can
also use the renormalized fields and couplings to write this lagrangian, so
that
all the terms keep same form
(as the theory is renormalizable in the generalized sense described above)
although
they are multiplied by renormalization $Z$ factors. The renormalized and the
bare
 fields (denoted with a 0 subscript) and gauge couplings are related as
follows:

\begin{eqnarray}
W_{0\mu}^a(x) =Z_3^{(a)1/2}W_{\mu}^a(x) ;
\omega_o^{\alpha}(x)=Z_{\omega}^{(\alpha)1/2}\omega^{\alpha}(x) ;
g_0^{(a)}=Z_g^{(a)}g^{(a)} ;
\xi_0^{(a)} =Z_3^{(a)}\xi^{(a)} 	  \\   \nonumber
c_0^a(x) =\widetilde Z_2^{(a)1/2}c^a(x) ;
\bar c_0^a(x)=\widetilde Z_2^{(a)1/2} \bar c^a(x) ;
B_0^a(x)=\widetilde Z_2^{(a)}B^a(x)
; v_0 =Z_v^{1/2}v
\end{eqnarray}
where $g^{(a)}=g$ for $a=1,2,3$ and $g^{(4)}=g'$ and from now on we
 use  indices between parenthesis as labels which are not summed.
Thanks to the gauge structure of the theory the first three $Z_3$ are equal.
Indeed, there are infinite relations between the bare and the renormalized
couplings
appearing in the chiral lagrangian. It is straightforward now to obtain a set
of
"renormalized" (anti)-BRS transformations leaving invariant the renormalized
lagrangian
written in terms of the renormalized fields and couplings:

\begin{eqnarray}
s_R[\omega^{\alpha}]=X^{(a)}L^{\alpha}_{Ra}c^a \;\;\;\;\;\; &\qquad&
\bar s_R[\omega^{\alpha}]=X^{(a)}L^{\alpha}_{Ra}\bar c^a \\  \nonumber
s_R[W^{\mu a}]=X^{(a)}D^{\mu a}_{Rc}c^c \;\;\; &\qquad&
\bar s_R[W^{\mu a}]=X^{(a)}D^{\mu a}_{Rc}\bar c^c \\  \nonumber
s_R[c^a]=-\frac{X^{(a)}}{2}f^a_{R\;bc}c^bc^c &\qquad&
\bar s_R[c^a]=-\frac{X^{(a)}}{2}f^a_{R\;bc}\bar c^bc^c \\ \nonumber
s_R[\bar c^a]=X^{(a)}\frac{B^a}{\sqrt{\xi^{(a)}}} \;\;\;\;\;\;\; &\qquad&
\bar s_R[\bar c^a]=-X^{(a)} \left( \frac{B^a}{\sqrt{\xi^{(a)}}} +
f^a_{R\;bc}\bar c^bc^c \right) \\ \nonumber
s_R[ B^a]=0 \;\;\;\;\;\;\;\;\;\;\;\;\;\;\;\;\;\;\; &\qquad&
\bar s_R[ B^a]=-X^{(a)} f^a_{R\;bc}\bar c^bB^c \\  \nonumber
\end{eqnarray}

 where $L^{\alpha}_{Ra}=Z_{\omega}^{(\alpha)-1/2}Z_3^{(a)1/2}L^{\alpha}_{\;a}$,
$f^a_{R\;bc}=Z^{(a)}_g Z_3^{(a)1/2} g f^a_{\;bc}$, and $X^{(a)}=\widetilde
Z_2^{(a)1/2} / Z_3^{(a)}$. One could think that the appearance of the $L$
factors, which are nonlinear in the GB fields, will  make the relation between
gauge bosons and GB extremely cumbersome, since this relation will be derived
from the BRS symmetry of the lagrangian. We will see that this is not the case.

 These "renormalized" symmetry of the quantum lagrangian will allow us in the
next section to apply the standard  functional methods
 to obtain the corresponding Ward-Slavnov-Taylor identities
 for renormalized Green functions which lead us to the ET.

\section{Ward-Slavnov-Taylor Identities}

Our aim is to obtain the relationship between $S$-matrix elements involving
 longitudinal gauge bosons $W_L$ and those elements where we have replaced
 the external $W_L$ by GB. To that end we will first obtain, from the
BRS invariance of the renormalized lagrangian, Ward-Slavnov-Taylor identities
that later will be translated, using the
 Lehmann-Symanzik-Zimmermann (LSZ) reduction formula, in relations
between $S$-matrix elements.

We start by remembering that the  generating functional for renormalized
 connected Green's functions $W_R(x_1,...,x_n)$  is given, in momentum
space, by the following definition:
\begin{equation}
W_R[J]=(2\pi)^4 \sum_{n=1} \int \prod_{i=1}^n \frac{d^4p_i}{(2\pi)^4}
\delta^4(\sum_i p_i)J_{i_1}(-p_1)...J_{i_n}(-p_n) W_{
R \; i_1,...,i_n}(p_1,...p_n)
\end{equation}
where $W_{R \; i_1,...,i_n}(p_1,...p_n)$ are renormalized Green functions.
We can now use	the BRS invariance of the lagrangian ($A$ stands for
any field appearing in the quantum lagrangian i.e. $A_i=\omega^{\alpha},
 W^a_{\mu}, c^a, \bar c^a, B^a$) to write:
\begin{equation}
\sum_i \int d^4x <s_R[A_i]>_J J_i(x)=0
\end{equation}
where we can write, in general, the BRS transformations as sums of linear or
nonlinear field products as:
\begin{equation}
<s_R[A_i]>_J=\sum_n s_{A_i}^{i_1...i_n} <A_{i_1}...A_{i_n}>_J
= \sum_n s_{A_i}^{i_1...i_n}
\frac{\delta^{(n)}W_R[J]}{\delta J_{i_1}...\delta J_{i_n}}
\end{equation}
Where, as usual, $J_i$ is the external current corresponding to the $A_i$
field.
In the GB case the above expression corresponds to the series expansion of the
nonlinear BRS transformation, for the rest of the fields there will appear
terms
with just one field, or in the case of $A_i=W^{\mu},c^a$ also terms with a two
field product. Therefore the BRS invariance condition can be written as:
\begin{equation}
I[J] = \sum_i \sum_n s_{A_i}^{i_1...i_n} \int
\frac{d^4q d^4k_1...d^4k_{n-1}}{(2\pi)^{4n}}
 \frac{\delta^{(n)}W_R[J]}{\delta J_{i_1}(q-k_1)...\delta J_{i_n}(k_{n-1})}
 J_i(-q) =0
\end{equation}

	It is now straightforward to obtain Ward-Slavnov-Taylor identities by
taking functional derivatives with respect to $J_i(p)$ at
$J=0$. Indeed, we are interested on identities involving the $B$ field, which
is nothing but the gauge fixing condition that intuitively identifies $W_L$ and
the GB.
Therefore we write:
\begin{equation}
 \left.\frac{\delta}{\delta J_{\bar c_{a_1}}(-k)}
\prod_{j=2}^{s}\frac{\delta}{\delta J_{B_{a_j}}(-k_j)}
\prod_{k=1}^{m}\frac{\delta}{\delta J_{A_k}(-p_k)} I[J] \right|_{J=0} =0
\end{equation}
Where we will impose that the currents $J_{A_k}$ are only
associated to physical $A_k$ fields. Taking just two functional derivatives
we obtain the following relation between the two leg Green functions
 (see \cite{DoPe} for details):
\begin{equation}
 \frac{X^{(b)}}{\sqrt{\xi^{(b)}}}W_{B^{b}l}(p)
=- X ^{(a)}D^{a}_{ R l}(p) W_{c^{a}\bar c^{b}}(p)
 \end{equation}
where:
\begin{equation}
D^{a}_{ R l}(p) = ip_{\mu}(1+\Delta_3(p^2))\delta_l^{W_{\mu a}}+
(L^{(0) \alpha}_{R a}+ \Delta_{2a}^{\alpha}(p^2))
\delta_l^{\omega^{\alpha}}
 \end{equation}
and we have used:
\begin{eqnarray}
ip_{\mu}\Delta_3(p^2)&=&f^a_{Rdc}W^{-1}_{c^{a}\bar
c^{b}}(p) \int \frac{d^4}{(2\pi)^4} W_{W^{\mu d}c^{c}\bar c^{b}}(p-q,q,p)\\
\nonumber \Delta_{2a}^{\alpha}(p^2)&=& L_{Ra}^{(1)\alpha
\beta}W^{-1}_{c^{c}\bar
c^{b}}(p) \int \frac{d^4}{(2\pi)^4} W_{\omega^{\beta}c^{c}\bar c^{b}}(p-q,q,p)
+
... \nonumber
\end{eqnarray}
Note three important features that were not present in the formal proof of
\cite{ChMKG}:
a) the renormalization $X$ factors, b) the $L_R^{(0)}$ term which is the term
coming only from the linear part of the GB BRS transformation, and that will
be, at
the end, the only remainder of the complicated realization of the symmetry at
lowest order in $g$ or $g'$, thus simplifying the relation between GB
and gauge bosons that one would expect naively from the nonlinear gauge fixing
condition; and c) the appearence of the $\Delta$ terms that were correctly
introduced
in \cite{Bagger} for the SM, and more recently in \cite{He} in the context of
Chiral Lagrangians. It is important to remark that these $\Delta_2$ and
$\Delta_3$
 terms are of higher order in $g$ or $g'$ than $ L^{(0)}_{R}$ and $1$,
 respectively, and therefore we will neglect them when using only the lowest
order
in the weak couplings.

In order to obtain the general expression, we have to notice from the BRS
transformations that we will  not get any contribution if $A_i=B$,
neither when $A_i=\omega,c$ because there are no $J_{\omega}$ nor $J_c$
derivatives. As the $A_k$ are physical, their polarization vectors will
cancel the derivative term in $s_R[W^a_{\mu}] = ik_\mu
c^{a}+\epsilon^a_{Rbc}W_{\mu b}c_c$ since $\epsilon \cdot k_\mu =0$. Therefore,
we
only have to take into account the contributions from $s_R[\bar c]$ and the
part which is
left from $s_R [W^a_{\mu}]$ that will be called "bilinear terms".
Thus we obtain:
\begin{equation}
 \frac{X^{(a_1)}}{\sqrt {\xi^{(a_1)}}}
W_{B_{a_1}B_{a_2}...B_{a_s}A_1...A_m}(k_1,...,k_s,p_1,...p_m) + \mbox{
bilinear terms} =0
\end{equation}
where $\sum_i k_i =-\sum_i p_i$. As the $a_1$ index is free we can drop the
factor
 $X/ \sqrt{\xi}$ which is irrelevant. However this is a relation between Green
functions,
and we have to apply the LSZ reduction formula to translate it to $S$-matrix
elements:
\begin{eqnarray}
  \left( \prod_{i=1}^{m}W_{A_iA_i}
(p_i) \right) \sum_{l_j} \left( \prod_{j=1}^{s}W_{B_{a_j}l_j}(k_j) \right)
S^{off-shell}_{l_1..l_s A_1...A_m}(k_1...k_s,p_1...p_s)  \\ \nonumber
 + \mbox{ bilinear terms}=0
\end{eqnarray}

 The next step in the LSZ procedure is to	multiply the above equation
 by the inverse $A_i$ propagators, and to set their momenta on-shell
, that is $p^2_i=m_{A_i}^2$, then  the "bilinear terms" cancel since they
contain a Green's function with at least
one off-shell momentum, and therefore they do not have the pole needed to
compensate for $W_{A_iA_i}^{-1}(p_1) \rightarrow 0$ when $p_1^2=m_{A_1}^2$.
 We can now use eq.12 to substitute the $B$ field two point functions and
thus we get:

\begin{equation}
\left. \sum_{l_j} \left(\prod_{j=1}^{s}\frac{\sqrt {\xi^{(a_j)}}}{X^{(a_j)}}
X^{(c_j)} W_{c^{c_j}\bar c^{a_j}}(k_j) D^{c_j}_{ R l_j}(k_j) \right)
S^{off-shell}_{l_1..l_s A_1...A_m}(k_1...k_s,p_1...p_s)
\right|_{p^2_i=m_{A_i}^2} =0
\end{equation}
The $\sqrt {\xi^{(a_j)}}/X^{a_j}$ factors are again irrelevant since we have
not
contracted the $a_j$ indices.
We still have to multiply by the ghost inverse two point
functions $W^{-1}_{c^{d_j}\bar c^{a_j}}(k_j)$ which are non-diagonal in
principle.
In so doing we see that the  $d_j$ indices are free again and we can
drop the other $X$ factors. We finally obtain:
\begin{equation}
\left. \sum_{l_1...l_r}\prod_{i=1}^{s}D^{a_i}_{R l_i}(p_i) S^{off-shell}
_{l_1..l_s,A_1..A_m}(p_1..p_r,k_1..k_m) \right|_{p^2_i=m_{A_i}^2} =0
\end{equation}

\section{The Equivalence Theorem}

The last step in the LSZ formulae is to set all the momenta on-shell,
but before that, we have to obtain the physical combinations out of
 the $W_{\mu}$ fields which appear in the $D_R$ operator. That is achieved
by means of a transformation  $\widetilde
W^a_{\mu}= R^{ab}W^b_{\mu}$, whose most general form is :
\begin{equation}
\left ( {\matrix{ \widetilde W_{\mu}^1 \cr
\widetilde W_{\mu}^2 \cr
\widetilde W_{\mu}^3 \cr
\widetilde W_{\mu}^4  }} \right ) =
\left ( {\matrix{ W_{\mu}^- \cr W_{\mu}^+ \cr Z^{phys}_{\mu} \cr A^{phys}_{\mu}
}}
\right) = \left ( {\matrix{ 1/\sqrt{2}&i/\sqrt{2}&0&0\cr
1/\sqrt{2}&-i/\sqrt{2}&0&0\cr
0&0&cos \theta & -sin \theta \cr
0&0&sin \theta' & cos \theta'  }} \right )
\left ( {\matrix{ W_{\mu}^1 \cr W_{\mu}^2 \cr W_{\mu}^3 \cr W_{\mu}^4
 }} \right)
\end{equation}
The propagators of these new renormalized fields present poles in the right
 values of the corresponding physical masses. According to those definitions
we also introduce: $\widetilde L^{(0)b}_{R\alpha}=L^{(0)a}_{R\alpha}
(R^{-1})^{ba}$ and $\widetilde \Delta_i$. Therefore we obtain:
\begin{equation}
\sum_{l_1...l_r}\prod_{i=1}^{s}\widetilde{D}^{a_i}_{R l_i}(p_i)
S _{l_1..l_s,A_1..A_m}(p_1..p_r,k_1..k_m) = 0
\end{equation}
where now
\begin{equation}
\widetilde{D}^{a}_{ R l}(p) = ip_{\mu}(1+\widetilde \Delta_3(M_{phys}^2))
\delta_l^{\widetilde{W}_{R \mu a}}+
(\widetilde{L}^{(0) \alpha}_{ Ra} + \widetilde
\Delta_{2a}^{\alpha}(M_{phys}^2))
\delta_l^{\omega^{\alpha}}
\end{equation}
Notice that the  $p_i$ momenta are on-shell for the massive physical vector
bosons, and since the $\Delta$ terms do not depend on the energy or external
momenta. From now on we will use amplitudes instead of	$S$ matrix elements as
it
is customary in $\chi PT$. As a matter of fact it is more convenient to  obtain
the ET from the following relation between amplitudes that we will write
symbolically as: \begin{eqnarray}
\left( \prod_{i=1}^{n}\epsilon_{(L)\mu_i}\right)
T(\widetilde W^{\mu_1}_{a_1},..., \widetilde W^{\mu_n}_{a_n};A) = \hspace {7cm}
 \\ \nonumber
 = \sum_{l=0}^{n} (-i)^l \left( \prod_{i=1}^{l}v_{\mu_i} \right) \left(
\prod_{j=l+1}^{n}K^{a_j}_{\alpha_j} \right) \bar T(\widetilde W^{\mu_1}_{a_1}
...\widetilde W^{\mu_l}_{a_l},\omega_{\alpha_{l+1}} ...\omega_{\alpha_n};A)
\end{eqnarray}

  Where we  have introduced
$v_\mu=\epsilon_{(L)\mu} -p_\mu / m \simeq O(M_{phys}/E)$, we have omitted
 the irrelevant indices, and we have defined:
\begin{equation}
K^{\alpha}_{Ra}=\frac{\widetilde{L}^{(0) \alpha}_{Ra} + \widetilde
\Delta_{2a}^{\alpha}(M_{phys}^{(a)2})}{M^{(a)}_{phys}(1+\widetilde
\Delta_3(M_{phys}^{(a)2}))}
\end{equation}
 which do not depend on the momenta, and $\bar T$
which is the sum over all the amplitudes with independent permutations of
fields
and indices for a given $l$ value. This relation was first obtained in
\cite{ChMKG} but without taking into account the $K$ factors (see \cite{DoPe}).

 In the proof by Chanowitz and Gaillard the next step  was
to  neglect at high energies the terms containing $v_\mu$ factors since
$v_\mu \simeq O(M_{phys}/E)$ and the amplitudes in the MSM satisfy the
unitarity bounds and do not grow with energy. Therefore they were able to drop
at high energies all terms in the RHS of eq.22 but the one with $l=0$
which is precisely that with all external $\widetilde W_L$ substituted by
GB, thus obtaining the ET.  However, in our case we are not allowed to so
since the amplitudes in $\chi PT$
are obtained as a truncated series in the energy and we cannot simply neglect
the
 terms containing $v_{\mu}$ factors, but we have to use power counting methods
 to obtain the leading orders. A similar problem arises when dealing with the
large $m_H$ limit of the SM \cite{Veltman}.

The $K$ factors in eq.22 include the renormalization effects
on the ET which also appear in the Chiral Lagrangian formalism, as it has been
recently
 shown in \cite{He} where the authors arrive, using a
different  quantization procedure, to similar results to us for $g'=0$ and the
GB parametrization $U=exp(i\sigma^a \omega^a/v)$. They concentrate in the
renormalization factors which correct the ET version without performing
the power counting analysis. (See also \cite{Grosse} for a general discussion)

 When dealing with $\chi PT$ we can expand the
amplitudes as Laurent series in $E/4\pi v$ up to a positive power $N$ by
fixing the maximum number of derivatives in the Lagrangian. However, these
amplitudes should satisfy the Low Energy Theorems (second reference in
\cite{Sikivie})
 in the  $M^2 \ll E^2$ regime, so that the energy negative powers can be
written as
$(M/E)^{k}$. Therefore:
\begin{equation}
 \bar T(\widetilde W^{\mu_1}_{a_1}
...\widetilde W^{\mu_l}_{a_l},\omega_{\alpha_{l+1}} ...\omega_{\alpha_n};A)
\simeq
\sum_{k=0}^{N} a_{l}^k \left(\frac{E}{4\pi v} \right) ^k + \sum_{k=1}^{\infty}
a_{l}^{-k} \left( \frac{M} {E}\right)^{k}
\end{equation}
	Notice that in order to simplify the analysis we will set momentarily $g'=0$
 and that we have omitted the field indices in $a^k_l$. The
coefficients in this formulae, which can contain logarithms of the energy and
therefore
 should be understood formally,  can be expanded perturbatively on
$g$, for instance: $a_{l}^h=a_{lL}^h(1+ O(g/4\pi))$ where $a_{lL}^h$ is the
lowest
order term in the $g$ expansion.
 In most renormalization schemes we have $M \simeq
M_{phys} (1+O(g/4\pi) )$ and therefore we can write
$K^a_{\alpha}\simeq K^{a(0)}_{\alpha}+ K^{a(1)}_{\alpha} (g/4\pi)+....$ where
now these coefficients are energy independent. Once we introduce these
expansions
 in eq.22, if we neglect the order $O(M/E)$ and $O(E/4\pi v)^{h+1}$ terms, we
obtain:
\begin{eqnarray}
 ( \prod_{i=1}^{n}\epsilon_{(L)\mu_i} )
T(\widetilde W^{\mu_1}_{a_1},..., \widetilde W^{\mu_n}_{a_n};A)  \simeq
\hspace{8cm} \\ \nonumber
\simeq \left(\prod_{j=1}^{n}K^{a_j (0)}_{\alpha_j} \right)
\sum_{k=0}^{h} (a_{0L}^{k}(1 + O(g/4\pi )))  \left( \frac{E}{4\pi v} \right)^k
+O\left( \frac{M}{E} \right)+O\left( \frac{E}{4\pi v} \right)^{h+1}
\end{eqnarray}
 which is the precise formulation of the ET in the Chiral Lagrangian formalism
(for the sake of brevity the indices $\alpha$ of $a_0$).

It is this expression the one which will allow us to analyze the
applicability of the ET in the $\chi PT$ description of the SMSBS, since  if we
want
 these approximations to make sense, we are only left with the following
applicability window:
\begin{eqnarray}
 M \ll E \ll 4\pi v \\ \nonumber
 g/4\pi \ll (E/4\pi v)^{h+1}
\end{eqnarray}

The first inequality was present in the original formulation of the ET, and it
comes from neglecting the $O(M/E)$ contributions. The second inequality is
characteristic of $\chi PT$ and is due to the fact that in the Chiral
Lagrangian
formalism  we always obtain the amplitudes as truncated series in the energy,
neglecting the $O(E/4\pi v)^{h+1}$ term. If we want to obtain sensible physical
predictions, definitely we cannot trust the chiral expansion beyond $E=4\pi v$
(Usually much before). It is then expected that the amplitude in eq.25
calculated
with and without the ET would yield identical results beyond that energy limit,
but
it will not have any physical meaning. The last constraint comes from
neglecting
the  $O(M/E)$ term while keeping at the same time the $O(E/4\pi v)^{h}$
contribution
, since we expect the former to be much smaller than the latter.

It is straightforward now to generalize the preceding results to $g' \neq 0$
because $g'
\ll g$ as well as $ M_Z ^{phys} \simeq M_W ^{phys} \simeq M^{(a)}$ for any $a$
(all the different masses are of the same order when counting energy powers).
We only have to take into account the lowest order of the $a$ coefficients in
the $g$ or $g'$ expansion so that
the same reasoning we had used when $g' = 0$  is still valid.

\section{Some numerical results}

In order to check the validity of the ET for chiral
lagrangians obtained in the previous section we have explicitly
 computed the tree level cross sections (so that all $Z$ factors are
 equal to one) up to four
derivatives obtained from the chiral lagrangian considered in the second
reference
in \cite{App} for the processes, $Z_L^0 Z_L^0 \rightarrow
 Z_L^0 Z_L^0$ and $W_L^+W_L^-
\rightarrow Z_L^0Z_L^0$  and $W_L^{\pm}Z_L^0
\rightarrow W_L^{\pm}Z_L^0$. We have used chiral  coordinates for the
parametrization of the coset space $S^3$, i.e. we have grouped the GB fields in
a $SU(2)$ matrix field as $U(x)=\exp (i\omega^a\sigma^a/v)$
and we have worked in the Landau gauge. The computation was done in two ways;
first we have calculated the amplitude for the corresponding gauge bosons and
then  we
have projected them into their longitudinal components. Once we had the
$S$ matrix elements we have computed the cross sections which we have
compared with those obtained using the ET, that is, only with external GB.
It is important to remark that the  ET as stated in eq.25  only allows us to
use the lowest order in $g$ or $g'$, and therefore, in this case, we have only
taken into account GB internal lines.
For the sake of definiteness we have considered two different models. The first
one corresponds to a selection of the four derivative couplings that reproduces
the low energy behaviour of the MSM. Thus the couplings become a function of
the
Higgs mass which is taken to be equal to $1 TeV$. In the second model we select
the values of the couplings so that they correspond to a QCD-like theory with
$N_C=3$ (in the large $N_c$ limit).

Some preliminary results obtained after our numerical computations are shown in
the figures. Figs.1 and 2 display the comparison between the
high  energy behavior of the GB and the longitudinal components of the gauge
bosons cross sections for the MSM and QCD models respectively. As it can be
seen, a perfect agreement is found  between both cross sections in the three
studied channels. However, as it was commented in the previous section,
this agreement does not mean at all that those cross sections reproduce
properly the underlying physics (the MSM or the QCD model in this case),
since we have been taking into account only up to four derivative terms in our
computations. That is the reason why the $4 \pi v$ energy upper bound is set in
eq.26 for the validity of the  standard $\chi PT$ computations. The same
happens
in the $\chi PT$ description of the low-energy pion interactions where it is
well known that the standard four derivatives computation only describes
properly
the experimental data for energies well below $4 \pi f_{\pi}$.

However, from the considerations done in the previous section about the
applicability window of the ET, we do not expect such a good agreement in the
low-energy
regime, let us say below $\simeq 1.5 TeV$, since higher orders in $g$ or $g'$
may be not neglegible.
This fact is confirmed in fig.3 and fig.4 where we plot the same processes
considered in fig.1 and fig.2 but concentrating the attention in the low-energy
region ($\sqrt{s} \leq 1.5TeV$). In this regime, we can see that the
cross sections	obtained using the ET (and therefore, at lowest order in the
weak couplings), do not reproduce well those coming from the complete tree
level
computations. However, the discrepancies are only due to higher $g$ and
$g'$ orders, since we have performed the same comparison in the $g,g'
\rightarrow 0$ limit, and the curves obtained with the two procedures overlap.
This check confirms the ET statement of eq.25, but it seems that the
corrections due to the weak couplings are too relevant in this low-energy
regime
to be neglected. One could think that
 the amplitudes obtained using the ET as in eq.25 would improve if we
add
the tree level diagrams to the next order in $g$ and $g'$ with four
external GB fields but also internal gauge bosons.
However, from  the proof of the ET that we
have sketched here we can see that in order to reproduce the next
order in the weak couplings
for the $W$-scattering amplitudes, not only should have we considered
the
diagrams with internal gauge boson lines, but also those terms in eq.22 with
$v_{\mu}$ factors, as well as contributions from the $\Delta$ terms from the
$K$ correction factors. In fact, we have checked numerically that this
naive approximation does not render any relevant improvement.

In fig.3 and fig.4 we observe that for the  $Z_L^0 Z_L^0 \rightarrow Z_L^0
Z_L^0$
and $W_L^+W_L^- \rightarrow Z_L^0Z_L^0$ the ET predictions for the scattering
cross sections are more or less accurate but this is not the case at all in the
$W_L^{\pm}Z_L^0 \rightarrow W_L^{\pm}Z_L^0$ channel for both models. Therefore,
the numerical results seem to confirm our expectations on the impossibility of
a sensible simultaneous application of the standard $\chi PT$ and the ET. In
spite of what our  plots for the MSM and the QCD model seem to suggest,
 the disagreement not only appears in the $W_L^{\pm}Z_L^0
\rightarrow W_L^{\pm}Z_L^0$ channel. We have also made computations with other
arbitrary coupling choices of the four derivative terms in the
chiral lagrangian and we have found disagreements between the direct
computation and the ET predictions in the $W_L^+W_L^-
\rightarrow Z_L^0Z_L^0$ channel.

\section{Discussion and conclusions}

	We want now to remark the general applicability features of the ET as it
is stated in eq.25. We have derived that expression inside the
Chiral Lagrangian formalism through the use of Ward-Slavnov-Taylor identities
derived from the BRS invariance of the quantum lagrangian of a gauged NLSM.
The fact that we have used a formalism which is covariant under
reparametrizations
of the GB allows us to apply the results to any coordinate choice, simply by
changing the scalar $f$ function in eq.4 which determines the actual form of
the
$R_{\xi}$-gauge. Since we have included the renormalization effects, eq.25 is
ready to be used beyond the tree level.

However, and due to the restrictions imposed in the energy by the applicability
 window of eq.26, we could still ask about the real utility of the ET, which
shows
 the high energy relation between the GB and the $W_L$'s $S$-matrix elements,
in the context of the  $\chi PT$
description of the SMSBS, which is nothing but a low-energy description of
the GB dynamics.

Let us give an example: for the important case of two longitudinal gauge bosons
elastic scattering, the ET applicability range obtained from eq.26 would be
$1.7 TeV << E << 4\pi v \simeq 3.1 TeV$, when we include in the lagrangian
the terms up to four derivatives. To confirm this fact we have carefully
compared
the result of
a direct computation of the longitudinal components of the gauge bosons
from the four derivative chiral lagrangian at the tree level with the
corresponding ET predictions and we have shown the results in the previous
section. As expected we have found that the ET, works properly as stated in
eq.25, that is, at lowest order in $g$ or $g'$. Unfortunately the
description thus obtained does not reproduce properly the physics below
$1.7TeV$, since higher corrections in the weak couplings become relevant.
Besides, it is well know
that at high energies such as $3TeV$ the $\chi PT$ calculations in the four
derivatives approximation  cannot be trusted in many cases.

It seems then that the second constraint in eq.26, which is a lower energy
limit,
is much stronger than the first, since it can exclude the energy region where
$\chi PT$ works better. Even more, the restrictions  that this constraint
produces on the energy applicability  window get more and more severe
when the calculation is done for higher loops, or what it is the same,
for higher derivative terms in the Chiral Lagrangian.

 It is important to remember that this discussion in terms of energy
expansions,
is due to the fact that at high energies the effective lagrangian does not
yield
a good unitary behavior for the truncated amplitudes.
 This fact does not allow us to simply neglect the $v_{\mu}=O(M/E)$
 factors when extracting the
leading energy term in eq.22, since the amplitudes can contain positive powers
of $E$. Nevertheless, there are non-perturbative methods to implement
unitary in the $\chi PT$ amplitudes so that at high energies they will never
grow
 with a power
of $E$, and therefore we are allowed to directly drop the terms with $v_{\mu}$
factors, thus obtaining:
\begin{equation}
 \left( \prod_{i=1}^{n}\epsilon_{(L)\mu_i}\right)
T(\widetilde W^{\mu_1}_{a_1},...,\widetilde W^{\mu_n}_{a_n};A)\simeq
\left(\prod_{j=1}^{n}K^{a_j}_{\alpha_j} \right)
T(\omega_{\alpha_1} ...\omega_{\alpha_n};A) + O(M/E)
\end{equation}
which is the usual formal statement of the ET. This unitarization procedures
include the use of Pad\'e approximants and dispersion
relations \cite{Pade}, large N-limit \cite{largen}, etc..., and they can
enlarge considerably the ET applicability range. These two methods have been
shown to work very well in hadron physics, where they are even appropriate
to deal with resonances, and are expected to do so
in the effective lagrangian description of the SMSBS.

To conclude we want to remark that there are three different ways
to apply the ET: First we find the case of a renormalizable theory
 as, for example, the MSM, whose amplitudes present a	good high energy
 behavior and thus we arrive to the ET as stated in eq.27 with just a lower
energy bound. As a matter of fact the only problem is the computation of the
GB amplitudes and the $K$ factors in the chosen renormalization scheme. The
second possible scenario is when we use standard $\chi
PT$ to describe the SMSBS since now the amplitudes are truncated series
in the energy. As we have already discussed, the precise statement of the
theorem
 is that of eq.25 although it is only applicable in the energy range of eq.26.
This version of the theorem is weaker, but in
this case the computation of the $K$ factors is not so hard as in the previous
one  since we only need to know the lowest order in their $g$ and $g'$
perturbative expansion. The fact that the ET only holds in the effective
lagrangian
formalism to the lowest order in $g$ had already been suggested in
\cite{Bagger}.
 However, the numerical results shown in the
previous section seem to indicate that probably there is no energy
applicability
window for this case.
Finally if we describe the SMSBS  by means of unitarized
$\chi PT$ the version of
the ET is again that of eq.27 without an upper energy applicability bound
but including the $K$ factors.

\section{Acknowledgements}

 This work has been supported in part by the Ministerio de Educaci\'on y
Ciencia (Spain)(CICYT AEN90-0034). We are specially grateful to M.J.Herrero
and C.P. Martin for their interesting suggestions. A.D. also
thanks the Gregorio del Amo foundation (Universidad Complutense de Madrid)  for
support and S. Dimopoulos and  the Department of Physics of the Stanford
University for their kind hospitality during the first part of this work.

\section{Figure Captions}

{\bf Figure 1.} We display the three channels $Z_L^0 Z_L^0 \rightarrow Z_L^0
Z_L^0$,$W_L^+W_L^-\rightarrow Z_L^0Z_L^0$	and $W_L^{\pm}Z_L^0
\rightarrow W_L^{\pm}Z_L^0$ in the high energy regime for the MSM.The
continuous
lines correspond to the cross sections with external GB and the dashed
lines those with external gauge boson longitudinal components

{\bf Figure 2.} Now we show the three channels
$Z_L^0 Z_L^0 \rightarrow Z_L^0 Z_L^0$,$W_L^+W_L^-
\rightarrow Z_L^0Z_L^0$  and $W_L^{\pm}Z_L^0
\rightarrow W_L^{\pm}Z_L^0$ in the high energy regime for the QCD-like model.
The continuous lines correspond to the cross sections with external GB and the
dashed
lines those with external gauge boson longitudinal components

{\bf Figure 3.} Here are shown the three channels $Z_L^0 Z_L^0
\rightarrow Z_L^0 Z_L^0$,$W_L^+W_L^-\rightarrow Z_L^0Z_L^0$  and
$W_L^{\pm}Z_L^0
\rightarrow W_L^{\pm}Z_L^0$ in the low energy regime for the MSM. The
continuous
lines correspond to the cross sections with external GB and the dashed
lines those with external gauge boson longitudinal components

{\bf Figure 4.}  Again we display the three channels
$Z_L^0 Z_L^0 \rightarrow Z_L^0 Z_L^0$,$W_L^+W_L^-
\rightarrow Z_L^0Z_L^0$  and $W_L^{\pm}Z_L^0
\rightarrow W_L^{\pm}Z_L^0$ but in the low energy regime for the QCD-like
model.
The continuous lines correspond to the cross sections with external GB and the
dashed
lines those with external gauge boson longitudinal components

\end{document}